\makeatletter
\@ifundefined{@parse@version@dash}{%
\def\@parse@version#1{\@parse@version@0#1}
\def\@parse@version@#1/#2/#3#4#5\@nil{%
\@parse@version@dash#1-#2-#3#4\@nil}
\def\@parse@version@dash#1-#2-#3#4#5\@nil{%
  \if\relax#2\relax\else#1\fi#2#3#4 }
}{}
\makeatother  

\documentclass[prb,twocolumn,superscriptaddress,showpacs,floatfix,longbibliography]{revtex4-2}
\usepackage{mathrsfs,braket}
\usepackage{amssymb, amsbsy, amsmath, latexsym, dsfont, array, layout, graphicx,mathrsfs,color,ulem,bm}
\usepackage[colorlinks=true,citecolor=blue,urlcolor=blue,linkcolor=blue]{hyperref}

\newcommand{\Stefano}[1]{{\color{black} #1}}

\begin{document}
\title{Topological aspects in nonlinear optical frequency conversion}

\author{Stefano Longhi}
\affiliation{%
Dipartimento di Fisica, Politecnico di Milano, Piazza L. da Vinci 32, I-20133 Milano, Italy
}%
\affiliation{%
IFISC (UIB-CSIC), Instituto de Fisica Interdisciplinar y Sistemas Complejos - Palma de Mallorca, Spain
}%

\date{\today}

\begin{abstract}
Nonlinear optical frequency conversion, observed more than half a century ago, is a corner stone in modern applications of nonlinear and quantum optics. It is well known that frequency conversion processes are constrained by conservation laws, such as momentum conservation that requires phase matching conditions for efficient conversion. However, conservation laws alone could not fully capture the features of nonlinear frequency conversion. Here it is shown that topology can provide additional constraints in nonlinear multi-frequency conversion processes. Unlike conservation laws, a topological constraint concerns with the conserved properties under continuous deformation, and can be regarded as a new indispensable degree of freedom to describe multi-frequency processes. We illustrate such a paradigm by considering sum frequency generation under a multi-frequency pump wave, showing that, akin topological phases in topological insulators, topological phase transitions can be observed in the frequency conversion process both at classical and quantum level.
\end{abstract}

\maketitle

\section{Intoduction}
 Since the first observation of optical harmonics more than half a century ago \cite{r1}, frequency conversion and wave mixing processes in
nonlinear optical media \cite{r2,r3,r4,r4b} have enabled the manipulation and control of the
electromagnetic radiation to a great extent,
with a variety of applications ranging from coherent harmonic generation \cite{r4,r4b,r5,r6,r7} to ultrafast optics and nonlinear spectroscopy \cite{r8,r9,r10},  quantum optics \cite{r11,r12,r13,r14,r15,r15b,r16}, nonlinear imaging and biological microscopy \cite{r17,r18}, to mention a few.\\
Modern nonlinear optics has borrowed many concepts from
quantum mechanics and condensed-matter physics, and in return, enriched the variety of
theoretical and experimental platforms where  quantum
phenomena can be studied (see e.g. \cite{r19,r20,r21,r21b,r22,r23,r24,r24b,r25,r26,r27} and references therein).
Prominent examples include the geometric (Berry) phase accompanying nonlinear
frequency mixing \cite{r20}, adiabatic processes in frequency conversion \cite{r19,r24,r24b}, 
and the design of
novel photonic structures which combine topological phases of light with appreciable nonlinear response \cite{r20}, thus extending to the nonlinear realm the 
recent developments in the area of topological photonics \cite{r28,r29,r30,r31,r32}.
Recently, it has been suggested that various
nonlinear optical effects can be described 
in a unified fashion by topological quantities involving the Berry connection and Berry curvature \cite{r33}.\\
Frequency conversion processes in nonlinear $\chi^{(2)}$ media, such as sum/difference frequency generation and parametric down-conversion, are constrained by conservation laws:  energy, flux, momentum and angular momentum of photons should be conserved during the nonlinear interaction \cite{r2,r3,r34,r35}. Such conservation laws are expressed by well-known conditions, such as the Manley-Rowe relations 
and the phase matching requirement for momentum conservation. However, conservation laws alone could not fully capture the properties of nonlinear frequency conversion. In this work we unravel that, akin to topological phases in condensed matter physics \cite{r45,r46b,r46c}, topology can provide additional constraints to nonlinear multi-frequency conversion processes, which can undergo topological phase transitions. Unlike conservation laws, topology concerns with the conserved properties under continuous deformation, and can be regarded as a new indispensable degree of freedom to describe nonlinear frequency conversion processes.\\

To unveil the topological aspects underlying frequency conversion, let us consider the process of sum frequency generation (SFG), where two input photons at frequencies
$\omega_1$ (signal wave) and $\omega_2$ (pump wave) annihilate while, simultaneously, one photon at frequency $\omega_3=\omega_1+\omega_2$ (SFG wave) is created under perfect phase matching in the nonlinear crystal. The process is quite simple when we deal with single-frequency fields, while topological features emerge when we consider multi-frequency waves. Let us assume that we inject one signal photon at frequency $\omega_1$ and a stream of $N_2$ and $N_2^{\prime}$ pump photons at slightly different frequencies $\omega_2$ and $\omega_{2}^{\prime}=\omega_2+ \Omega$, respectively [Fig.1(a)]. Clearly, the signal photon can annihilate with one pump photon of either frequency $\omega_2$ or $\omega_{2}^{\prime}$, so that the frequency of the SFG photon can be either $\omega_{3}=\omega_1+\omega_2$ or $\omega_{3}^{\prime}=\omega_1+ \omega_{2}^{\prime}=\omega_3+ \Omega$ with probabilities $N_2/(N_2+N_2^{\prime})$ and $N_2^{\prime}/(N_2+N_2^{\prime})$, respectively. In repeated measurements, on average the frequency of the SFG photon is thus $\langle \omega_3 \rangle=\omega_3+ \nu \Omega$, with $\nu=N_2^{\prime}/(N_2+N_2^{\prime})$. \Stefano{Clearly, $\nu$ is not quantized, i.e. it not an integer number, and could be any real number depending on the values of $N_2$ and $N_2^{\prime}$. However,} this result holds for a short interaction length $z$:
\begin{figure}[!]
	\includegraphics[width=0.98\linewidth]{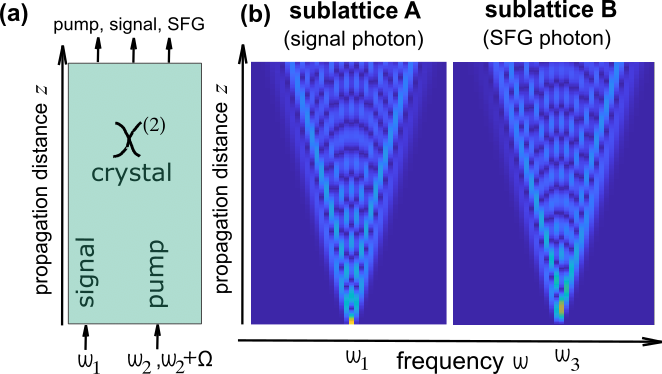}
\caption{\textbf{Multi-frequency SFG.} (a) A weak monochromatic signal wave at frequency $\omega_1$ interacts with a strong bi-chromatic pump wave, at frequencies $\omega_2$ and $\omega_2^{\prime}=\omega_2+ \Omega$, in a nonlinear crystal to generate a SFG wave. (b) The signal and SFG photons describe in tandem a quantum walk in a synthetic binary lattice in frequency space. The probability distribution of the photon frequency, depicted on a pseudocolor map, broadens as the interaction length $z$ increases. The topology of the synthetic lattice provides a constraint on the mean frequency of signal and SFG photons.}
\end{figure}
further interaction in the nonlinear crystal makes it possible the backward process, i.e. the newly generated SFG photon can annihilate and generate a pair of signal and pump photons. Energy conservation imposes that the frequency of the created signal photon should belong to the set $\{
 \omega_1 ,\;  \omega_1-\Omega, \; \omega_1+\Omega \}$.
Such a newly created signal photon can then annihilate with one pump photon to generate a SFG photon at a frequency that must belong to the set $\{ \omega_3-\Omega, \omega_3, \omega_3+\Omega, \omega_3+2 \Omega \}$ for energy conservation. This reasoning can be iterated and the frequency of both signal and SFG photons basically undergo a diffusion process in frequency space. Hence, as the interaction length $z$ in the nonlinear crystal increases, we have an evolving  probability distribution for the frequency of the created SFG photon. Energy conservation requires that such a frequency should belong to the set $\omega_3+n \Omega$ ($ n$ integer), but does not pose any constraint about the mean value $\langle \omega_3 \rangle$ of such a distribution, which in principle could take any value $\omega_3+ \nu \Omega$ with $\nu$ a real number. Here topology comes into play: as we show in this work, in the multi-frequency conversion process the signal and SFG photons describe in tandem a quantum walk on a topological lattice in synthetic (frequency) space [Fig.1(b)], resulting in the quantization of $\nu$ for long interaction lengths. Specifically, the integer $\nu$ turns out to be a topological invariant (winding number) associated to the synthetic lattice and determined by the multi-frequency properties of the injected strong pump wave. This is the main message of this work, which is developed and presented with the due mathematical details in the next sections.\\

\section{Topological signature in sum-frequency generation with a multi-frequency periodic pump wave}
\subsection{Classical analysis} The quantization of $\nu$ can be readily proved in the framework of a classical analysis of three-wave frequency mixing in a non-linear $\chi^{(2)}$ crystal. In the plane-wave approximation, the electric field propagating along the longitudinal $z$ direction of the crystal can be written as 
\[ \mathcal{E}(z,t)=\frac{1}{2} \left\{ \sum_{l=1}^3 \sqrt{\frac{2 \hbar \omega_l}{ \epsilon_0 c_0 n_l}}\psi_l \exp(-i \omega_l t+ik_l z) +c.c.\right \}, \] 
where $\omega_1$, $\omega_2$ and $\omega_3$ are the carrier frequencies of signal, pump and SFG waves, respectively, $k_l=(\omega_l/c_0)n_l$ are the wave numbers and $n_l=n(\omega_l)$ the (linear) refractive indices. Under perfect phase matching, the three coupled equations governing the evolution of the field envelopes $\psi_l(z,t)$ read (see e.g. \cite{r2,r3,r14,r24,r34,r34b})

\begin{eqnarray}
i \left(  \frac{\partial}{\partial z}  + \frac{1}{v_{g1,2}} \frac{\partial}{\partial t}  \right) \psi_{1,2} & = & - \sigma \psi_3 \psi_{2,1}^* \\
i \left(  \frac{\partial}{\partial z}  + \frac{1}{v_{g3}} \frac{\partial}{\partial t} \right) \psi_{3}  & = & - \sigma \psi_1 \psi_2,
\end{eqnarray}
where $\sigma \equiv [d_{e}/(n_1 n_2 n_3)]  \sqrt{2 \hbar k_1k_2k_3/ \epsilon_0}$, $d_{e}$ is the effective nonlinear interaction coefficient, and $v_{gl}=1/(dk/d \omega)_{\omega_l}$ is the group velocity at carrier frequency $\omega_l$. In the above equations, the field envelopes have been normalized such that $|\psi_l|^2$ is the photon flux of the e.m. wave at frequency $\omega_l$. 
As usual in problems of sum and difference frequency generation \cite{r2,r19,r24,r36}, we assume that the crystal is excited by a strong pump field, not necessarily monochromatic, and by a monochromatic weak signal at frequency $\omega_1$. In the undepleted pump approximation and after letting $\xi=z$ and $\eta=t-z/v_{g3}$, one has $\psi_2(\xi, \eta) \simeq \psi_2(\xi=0, \eta)$, and Eqs.(1,2) reduce to the linear two-level equations
\begin{eqnarray}
i \frac{\partial \psi_1} {\partial \xi} & = & i \left(\frac{1}{v_{g2}}-\frac{1}{v_{g1}} \right) \frac{ \partial  \psi_1}{\partial \eta}+h(\eta) \psi_3 \\
i \frac{\partial \psi_3} {\partial \xi} & = &  i \left( \frac{1}{v_{g2}}-\frac{1}{v_{g3}} \right) \frac{ \partial \psi_3}{\partial \eta} +h^*(\eta) \psi_1
\end{eqnarray}
 where $h(\eta) \equiv - \sigma \psi_2^*(\xi=0, \eta)$ describes the temporal shape of the injected strong pump wave. As shown in Appendix A, for a sufficiently spectrally-narrow pump wave the group velocity mismatch terms can be neglected, so that Eqs.(3,4) can be readily integrated with the initial condition $\psi_1(\xi=0,\eta)=1$ and $\psi_3(\xi=0,\eta)=0$, yielding 
 \[ \psi_1(\xi,\eta)= \cos[\Delta(\eta) \xi] \] 
 \[ \psi_3(\xi,\eta)=-i \sin [ \Delta (\eta) \xi ] \exp[-i \varphi(\eta)], \]
  where we have set $h(\eta) \equiv \Delta (\eta) \exp[i \varphi (\eta)]$, \Stefano{i.e. $\Delta(	\eta)$ and $\varphi(\eta)$ are the amplitude and phase of the normalized pump wave}. Let us now assume that $h(\eta)$ is periodic with period $T=2 \pi/ \Omega$, i.e. that the pump wave carries a stream of photons at frequencies $\omega_2+n \Omega$, and let us set $k=\Omega \eta$. Correspondingly, the signal and SFG wave $\psi_
{1,3}(\xi, k)$ are periodic with respect to $k$ with period $2 \pi$ and can be thus written as a Fourier series, $\psi_{1,3}(\xi,k)=\sum_l (a_l,b_l) (\xi) \exp(-il k)$ with $\xi$-dependent amplitudes $a_l (\xi),b_l(\xi)$. At the propagation distance $\xi$, the mean of the frequency of the signal wave, given by 
$\langle \omega_1 \rangle= \omega_1+ \sum_l l \Omega  |a_l(\xi)|^2 / \sum_l |a_l(\xi)|^2$, reads $\langle \omega_1 \rangle=\omega_1$, whereas
the mean frequency of the SFG wave, given by $\langle \omega_3 \rangle= \omega_3+ \sum_l l \Omega  |b_l(\xi)|^2 / \sum_l |b_l(\xi)|^2$, can be written as 
$  \langle \omega_3 \rangle= \omega_3+ \nu \Omega$, where we have set (technical details are given in Appendix A)
 \begin{equation}
 \nu= \frac{\int_{-\pi}^{\pi} dk \sin^2 [\Delta(k) \xi] \frac{\partial \varphi}{\partial k}} {\int_{-\pi}^{\pi} dk \sin^2 [\Delta(k) \xi] }
 \end{equation}
If we assume that $\Delta(k) \neq 0$ for any $k$, i.e. that the pump wave is non-vanishing at any time instant, for long interaction lengths $\xi$ we can set $\sin^2 [\Delta(k) \xi] \simeq 1/2$ in Eq.(5), yielding $\nu \simeq (1/2 \pi) \int_{-\pi}^{\pi} dk ( d \varphi / dk) \equiv \nu_{\infty}$. This relation clearly shows that the index $\nu$ is quantized and equals the phase spanned by the pump wave in one oscillation cycle, normalized to $2 \pi$. For example, for an injected bichromatic pump at frequencies $\omega_2$ and $\omega_2^{\prime}=\omega_2+\Omega$, $h(k)=h_0+h_1 \exp(ik)$ and thus $\nu=0$ for $|h_0|>|h_1|$ and $\nu=1$ for $|h_0|<|h_1|$, the case $|h_0|=|h_1|$ corresponding to a topological phase transition.\\
 To illustrate the quantization of $\nu$ in a realistic setting, let us consider SFG in a periodically-poled lithium niobate (PPLN) crystal with a strong pump at the wavelength $\lambda_2 = 810$ nm and  a weak signal at $\lambda_1=1.55 \; \mu$m. The SFG wave corresponds to
$\lambda_3 =532$ nm. We assume extraordinary wave propagation, with a nonlinear coefficient $d_{33} \simeq 27$ pm/V. Phase matching is realized by a
first-order QPM grating ($7.38 \; \mu$m period), so that $d_{e}=(2/ \pi) d_{33}$ \cite{r38}.
Figure 2 shows the behavior of the index $\nu$ versus propagation distance $z$ in the crystal for a bichromatic pump wave
with a frequency offset $\Omega=2 \pi \times 1$ GHz and with two different values of the ratio $h_1/h_0=\sqrt{(I_1/I_0)}$ between the two harmonic pump amplitudes. The simulations take into account group velocity mismatch, as calculated using Sellmeier  equations for $n(\omega)$ \cite{r39}. The figure clearly illustrates the asymptotic quantization of $\nu$ for long interaction lengths and the topological phase transition as the ratio of pump intensities $I_1/I_0$ varies from below to above one.  

\begin{figure}[h]
	\includegraphics[width=0.98\linewidth]{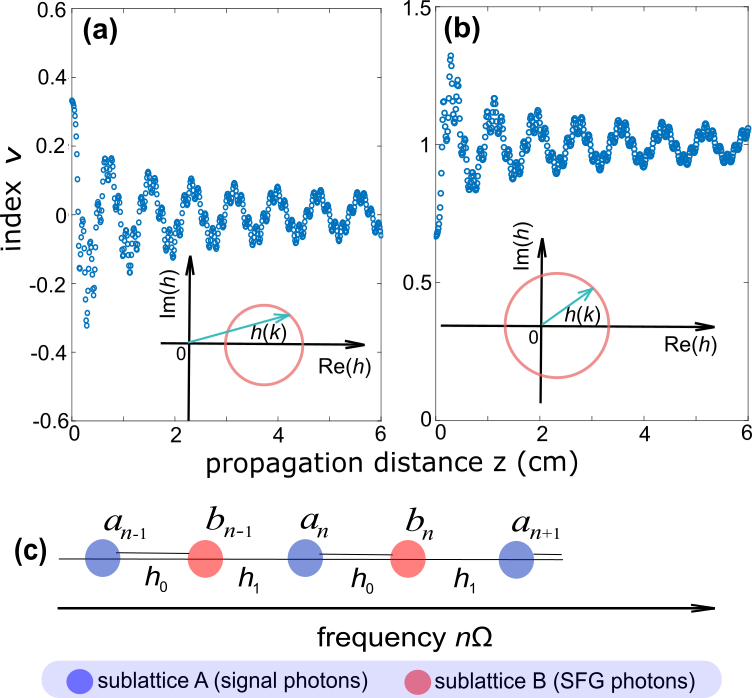}
\caption{\textbf{Quantization of index $\nu$.} (a,b) Behavior of the index $\nu$ versus propagation distance in the process of SFG in a PPLN crystal. The pump wave is bichromatic with pump intensities $I_0$ and $I_1$ at frequencies $\omega_p$ and $\omega_p+\Omega$. In (a) $I_0=800$ MW/cm$^2$, $I_1=400$ MW/cm$^2$; in (b) 
$I_0=400$ MW/cm$^2$, $I_1=800$ MW/cm$^2$. The insets show the behavior of $h(k)=h_0+h_1 \exp(ik)$ in complex plane, parametrized in the scaled time $k=\eta \Omega$. Parameter values are given in the text. \Stefano
{(c) Synthetic SSH lattice in frequency space along which the signal and SFG photons undergo a quantum walk in tandem.}}
\end{figure}
\subsection{Quantum analysis} The quantization of the index $\nu$ predicted by the classical analysis can be at best captured in the second-quantization framework of SFG \cite{r36,r40,r41,r42,r43,r44}. Here, the signal and SFG photons undergo in tandem a quantum walk on a synthetic lattice with nontrivial topology in frequency space, the index $\nu$ corresponding to a topological invariant of the lattice. The second-quantization analysis shows that the topological origin of $\nu$-quantization holds for an arbitrary non-classical state of the injected signal wave, i.e. not necessarily for classical (coherent) states. As in the classical analysis, we assume a multi-frequency pump with frequencies $\omega_2+n \Omega$,
centered at around the carrier $\omega_2,$ and neglect group-velocity mismatch effects. The second-quantization Hamiltonian of the photon field then reads \cite{r36,r40} 
\[ \hat{H}=\hat{H}_0+\hat{H}_I,\]
 where 
 \[ \hat{H}_0=\sum_n \hbar (\omega_1+n \Omega) \hat{a}^{\dag}_n \hat{a}_n+ \sum_n \hbar (\omega_3+n \Omega) \hat{b}^{\dag}_n \hat{b}_n+  \]
 \[+  \sum_n \hbar (\omega_2+n \Omega) \hat{c}^{\dag}_n \hat{c}_n\]
  is the Hamiltonian of the free field, 
  \[ \hat{H}_I=- \hbar \sigma v_{g} \sum_{n,l}(\hat{b}_{n+l} \hat{a}^{\dag}_l \hat{c}^{\dag}_n+H.c.) \]
   is the interaction Hamiltonian, $\hat{a}_n$, 
$\hat{b}_n$ and $\hat{c}_n$ are the bosonic annihilation operators of photon modes at frequencies $\omega_{1}+n \Omega$, $\omega_{3}+n \Omega$ and $\omega_{2}+n \Omega$, respectively. Assuming a strong and classical pump wave, the operators $\hat{c}_n$ can be considered as c-numbers \cite{r36}, and the Heisenberg equations of motion of the destruction operators $\hat{a}_n$, $\hat{b}_n$, after the transformation $\hat{a}_n \rightarrow \hat{a}_n \exp[-i(\omega_1+n \Omega) t]$, $\hat{b}_n \rightarrow \hat{b}_n \exp[-i(\omega_3+n \Omega) t]$, read (see Appendix B for details)
\begin{eqnarray}
i \frac{d \hat{a}_n}{dt}=- \sigma v_{g} \sum_l C_l^* \hat{b}_{n+l} , \; i \frac{d \hat{b}_n}{dt}=- \sigma v_{g} \sum_l C_l \hat{a}_{n-l} \;\;\;\;
\end{eqnarray}
where $C_n=\langle c_n \rangle$ and the interaction time $t$ is related to the interaction length $\xi$ by the relation $t=\xi / v_{g}$. Equation (6) indicates that the signal and SFG photons undergo in tandem a continuous-time quantum walk on the sublattices A and B of a one-dimensional (1D) lattice with chiral symmetry and long-range hopping amplitudes $\sigma v_{g}C_l^*$, which provides an extension of the famous Su-Schrieffer-Heeger (SSH) 1D topological insulator \cite{r45,r46}. Note the the c-numbers $C_l$ are basically the Fourier amplitudes of the classical strong pump waveform, namely $\psi_2(k)=\sum_l C_l \exp(-i k l)$, with $k= \Omega \eta$. After letting $\hat{\psi_1}(k,t)=\sum_n \hat{a}_n \exp(-ikn)$ and $\hat{\psi_3}(k,t)=\sum_n \hat{b}_n \exp-(ikn)$, the evolution equations for the operators  $\hat{\psi}_{1,3}(k,t)$ read $i  (d/dt) (\hat{\psi}_1, \hat{\psi}_3)^T=v_{g}H(k) (\hat{\psi}_1, \hat{\psi}_3)^T$ with matrix Hamiltonian 
\begin{eqnarray}
H(k)& = & 
\left( 
\begin{array}{cc}
0 & h(k) \\
h^*(k) & 0
\end{array}
\right) \nonumber \\
& = & \Delta(k)\left\{  \cos [ \varphi(k)] \sigma_x-\sin [ \varphi(k)] \sigma_y \right\} \;
\end{eqnarray}
where we have set
\[ h(k)=-\sigma \psi_2^*(k) \equiv \Delta(k) \exp[i \varphi(k)] \]
  and $\sigma_{x,y}$ are the Pauli matrices. Note that the Heisenberg equations for the $\hat{\psi}_{13}$ operators are analogous to the classical ones [Eqs.(3) and (4)] with $v_{g1}=v_{g2}=v_{g3}=v_g$ after the substitution $t \rightarrow z/v_{g}$ and considering $\hat{\psi}_{1,3}(k,t)$ as $c$-numbers.   
 
   Let us assume that the crystal is excited with a monochromatic signal field at frequency $\omega_1$ in an arbitrary quantum state, given by a superposition of Fock states   $|\psi(0) \rangle =\sum_{l=1}^{\infty} ( \alpha_l 
/ \sqrt{l !}) \hat{a}^{\dag l}_0 |0 \rangle$ with arbitrary amplitudes $\alpha_l$ and $\sum_l | \alpha_l|^2=1$. Note that excitation with a single-photon Fock state corresponds to $\alpha_l=\delta_{l,1}$, whereas excitation with a classical field (a coherent state) corresponds to a Poisson distribution $\alpha_l=\alpha^l \exp(- |\alpha|^2/2) / \sqrt{l!}$, with $\alpha= \psi_1(0)$. After a propagation distance $\xi=v_{g} t$, the mean value of the frequency of the signal and SFG photon fields can be readily calculated and read (details are given in Appendix B)
\[ \langle \omega_1 \rangle = \omega_1 \;, \;\; 
 \langle \omega_3 \rangle = \omega_3+ \nu \Omega, \]
  where 
 the value of $\nu$ is the same as the one obtained from the classical analysis [Eq.(5)], regardless of the initial state $|\psi(0) \rangle$ of the signal photon field. 
 
 \Stefano{\subsection{Frequency conversion and winding number}}
 \Stefano{The main result, that unravels the topological aspects in the SFG process, is that for long interaction lengths $\xi$ the index $\nu$ converges to the topological invariant (winding number) $\nu_{\infty}$ of the 1D gapped topological insulator. For example, if we assume a bichromatic pump as in the simulations of Fig.2, corresponding to $h(k)=h_0+h_1 \exp(ik)$, the signal and SFG photons undergo a quantum walk on a synthetic SSH lattice in frequency space with alternating hopping amplitudes $h_0$ and $h_1$ [see Fig.2(c)], the two sublattices A and B corresponding to the various frequency components $\omega_{1,3}+ n \Omega$ of the two fields. The topological invariant of a 1D gapped topological insulator with chiral symmetry is provided by the Zak phase $\gamma_{\pm}$ of the two lattice bands, given by \cite{r45}
 \[
 \gamma_{\pm}=i \int_{-\pi}^{\pi} dk \langle \mathbf{u}_{\pm} | \frac{\partial}{\partial k} \mathbf{u}_{\pm} \rangle=\frac{1}{2} \int_{-\pi}^{\pi} dk \frac{\partial \varphi}{\partial k}=\pi \nu_{\infty}
 \]
where 
\[
\mathbf{u}_{\pm}= \frac{1}{\sqrt{2}} \left(
\begin{array}{c}
1 \\
\pm \exp[-i \varphi(k)]
\end{array}
\right)
\]
 are the two eigenstates of the Bloch Hamiltonian $H(k)$ [Eq.(7)] corresponding to the eigen-energies $\pm |h(k)|$, and
 \[
 \nu_{\infty}= \frac{1}{2 \pi} \int_{-\pi}^{\pi} dk \frac{\partial \varphi}{\partial k}
 \]
 is the winding number. Note that the Zak phase in the two bands takes and same value, related to the winding number $\nu_{\infty}$, and that $\nu_{\infty}$ is the asymptotic value of $\nu (\xi)$ [Eq.(5)] as $ \xi \rightarrow \infty$. The
quantization of $\nu$ as $ \xi \rightarrow \infty$, such as the one observed in Fig.2(a,b), can be explained in terms of the asymptotic quantization of the mean displacement that the signal and SFG photons undergo in the tandem quantum walk in the synthetic frequency space. In fact, as shown in previous works \cite{r47,r48,r49,r50,r51,r52} for a gapped 1D topological insulator such a mean displacement is asymptotically quantized and equals the winding number $\nu_{\infty}$ of the topological lattice.} According to the bulk-boundary correspondence \cite{r45,r46,r53}, $|\nu_{\infty}|$ measures the number of topologically-protected zero-energy edge states, and the quantum walk provides a bulk probing method to measure $|\nu|$ \cite{r47}.\\
\\
\section{Topological signatures under a multifrequency aperiodic pump} The previous analysis can be extended to the case where the envelope $\psi_2(\eta)$ of the strong pump wave  is aperiodic in time and given by the superposition of $N$ mutually-incommensurate frequencies $\Omega_1$, $\Omega_2$,..., $\Omega_N$. In this case, the signal and SFG photons undergo a quantum walk on a high-dimensional synthetic lattice in frequency space \cite{r54}, which can display nontrivial topological features.\\
 Let us consider the simplest case of $N=2$ incommensurate frequencies $\Omega_1$ and $\Omega_2$, and let $k_1=\Omega_1 \eta$ and $k_2=\Omega_2 \eta$. The temporal pump waveform $\psi_2(\eta)$ can be considered as a periodic function of the two variables $k_1$, $k_2$ and expanded in double Fourier series as 
 \[  \psi_2(k_1,k_2)=\sum_{n,m} C_{n,m} \exp(-ik_1n-ik_2m). \]  
In the classical model of SFG, neglecting group velocity mismatch effects and assuming a monochromatic injected signal field at the entrance of the crystal, the solution to Eqs.(3) and (4) is given by 
\[ \psi_1(\xi,k_1,k_2)= \cos[\Delta(k_1,k_2) \xi] \]
\[ \psi_3(\xi,k_1,k_2)=-i \sin [ \Delta (k_1,k_2) \xi ] \exp[-i \varphi(k_1,k_2)], \]
 where $h(k_1,k_2) \equiv - \sigma \psi_2^*(k_1,k_2)$ is written in terms of amplitude and phase as  $h(k_1,k_2) \equiv \Delta (k_1,k_2) \exp[i \varphi (k_1,k_2)]$.
 The mean frequencies of the signal and SFG photons read $\langle \omega_1 \rangle =\omega_1$ and $\langle \omega_3 \rangle =\omega_3+\nu_1 \Omega_1+\nu_2 \Omega_2$, where we have set (technical details are given in Appendix C)
\begin{equation}
 \nu_{
1,2}= \frac{\iint_{-\pi}^{\pi} dk_1 dk_2 \sin^2 [\Delta(k_1,k_2) \xi] \left( \frac{\partial \varphi}{\partial k_{1,2}} \right)} {\iint_{-\pi}^{\pi} dk_1 dk_2 \sin^2 [\Delta(k_1,k_2) \xi] }
\end{equation}
Assuming that $\Delta(k_1,k_2) \neq 0$, i.e. that the pump wave $\psi_2(\eta)$ does not vanish for any time instant $\eta$, for long enough propagation distances we may set $\sin^2 [ \Delta(k_1,k_2) \xi] \simeq 1/2$ in Eq.(8), yielding 
\begin{equation}
\nu_{1,2}=(1/2 \pi) \int_{-\pi}^{`\pi} dk_{1,2} ( \partial \varphi / \partial k_{1,2}).
\end{equation}
\begin{figure}[h]
	\includegraphics[width=0.9\linewidth]{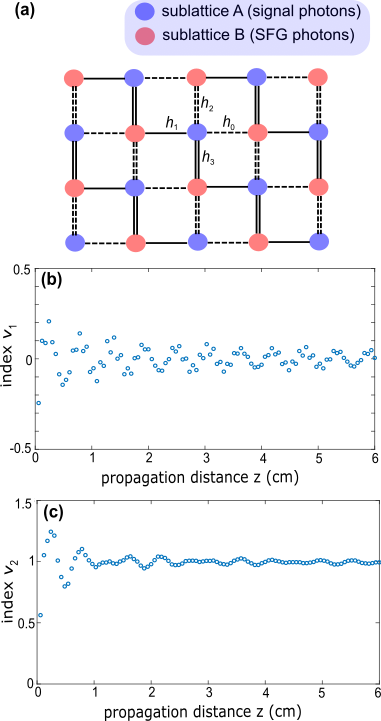}
\caption{\textbf{Topological indices with a multifrequency aperiodic pump.} \Stefano{(a) Schematic of the 2D synthetic topological lattice in frequency domain corresponding to a pump wave comprising four harmonic terms at frequencies $\omega_2$, $\omega_2 -\Omega_1$, $\omega_2+\Omega_2$, and $\omega_2-\Omega_1-\Omega_2$ with amplitudes $h_0$, $h_1$, $h_2$ and $h_3$, respectively. }(b,c) Behavior of the indices $\nu_{1,2}$ versus propagation distance in a PPLN crystal; parameter values  are given in the text.}
\end{figure}
 The values of $\nu_{1,2}$ turn out to be independent of $k_{1,2}$ and are integer indices (winding numbers). The same result holds in the second-quantization framework of SFG, and the indices $\nu_
{1,2}$ correspond to the topological numbers of a weak 2D topological insulator, along which correlated signal and SFG photons undergo a tandem quantum walk. In fact, in the second-quantization framework the photon fields of signal and idler waves are given in terms of the bosonic operators $\hat{a}_{n,m}(t)$, $\hat{a}^{\dag}_{n,m}(t)$  and $\hat{b}_{n,m}(t)$, $\hat{b}^{\dag}_{n,m}(t)$, respectively, that annihilate and create photons at frequencies $\omega_1+n \Omega_1+ m \Omega_2$ and $\omega_3+n \Omega_1+ m \Omega_2$, respectively. In the limit of a strong classical pump wave, the Heisenberg equations of motion of the destruction operators are a 2D extension of Eq.(6)  (see Sec.2 of Appendix C), and formally describe a quantum walk in two sublattices A and B of a 2D lattice in synthetic frequency space \cite{r54} with an Hamiltonian in Bloch space given by Eq.(7), with the replacement $k \rightarrow (k_1,k_2)$. Such a lattice is a 2D extension of the SSH model \cite{r55,r56,r57,r58} and provides an important example of a 2D weak topological
insulator \cite{r57,r58} sustaining flat-band edge states \cite{r55,r56}.
 The Hamiltonian $H(k)$ shows chiral and parity-time reversal symmetries, namely $H (k) \sigma_z=- \sigma_z H(k)$ and $\mathcal{PT}H(k)=H(k) \mathcal{PT}$, where $\mathcal{P}=\sigma_x$ and $\mathcal{T}= \mathcal{K}$ (complex conjugation) are the parity and time reversal operators. Moreover, provided that the Fourier coefficients $C_{n,m}$ of the pump wave are real, $H(k)$ also displays inversion symmetry \cite{r59}, namely $H(-k) \mathcal{P}=\mathcal{P} H(k)$. For such a 2D lattice, the Berry curvature identically vanishes and the topological phases can be identified by the strong $\mathbb{Z}_2$ index $\nu_0$ and by two weak $\mathbb{Z}_2$ indices $\bar{\nu}_{1,2}$  \cite{r57,r58}, or equivalently by the  vectorized Zak phase in 2D \cite{r60,r61,r62}.  Technical details are given in Sec.3 of Appendix C. The strong index $\nu_0=0$ corresponds to the insulating (i.e. gapped) phase, which is equivalent to the condition $\psi_2(\eta) \neq 0$, whereas the 2D vectorized Zak phase can be mapped  into the quantized indices $\nu_{1,2}$ [Eq.(9) mod 2]. Therefore, the mean frequency $\langle \omega_3 \rangle$ of the SFG wave in the gapped phase $\psi_2(\eta) \neq 0$ is constrained by topological properties of the 2D weak topological insulator.\\ 
As an illustrative example, let us assume 
\begin{eqnarray}
h(k_1,k_2) & = & h_0+h_1 \exp(-ik_1)+h_2 \exp(ik_2)+ \nonumber \\
 & + & h_3 \exp(-ik_1-ik_2),
\end{eqnarray} 
which corresponds to a pump envelope $\psi_2(\eta)= -(1/ \sigma) h^*(k_1,k_2)$ comprising the four frequencies $\omega_2$, $\omega_2 -\Omega_1$, $\omega_2+\Omega_2$, and $\omega_2-\Omega_1-\Omega_2$ with amplitudes $h_0$, $h_1$, $h_2$ and $h_3$, respectively. \Stefano{The topological 2D synthetic lattice, along which the SFG and signal photons undergo a tandem quantum walk, is shown in Fig.3(a). Note that the amplitudes of the four pump waves correspond to the hopping amplitudes in the synthetic 2D lattice}. The value of the strong  topological  index $\nu_0$ can be computed from the parity eigenvalue of the Bloch eigenstates at the four time-reversal invariant momenta $(k_1,k_2)= \pi (n_1,n_2)$  ($n_{1,2}=0,1$) [see Eq.(C16) in Appendix C], and reads
\begin{eqnarray}
(-1)^{\nu_o} & = & {\rm{sign}} \left\{ (h_0+h_1+h_2+h_3)(h_0-h_1+h_2-h_3)   \right\} \nonumber \\
& \times & {\rm{sign}} \left\{ (h_0+h_1-h_2-h_3)(h_0-h_1-h_2+h_3)  \right\}. \nonumber
\end{eqnarray}
In the insulating phase, i.e. for $\nu_0=0$, the winding numbers $\nu_{1,2}$ can be calculated from Eq.(C21) of Appendix C along the lines $k_{2,1}=0$, i.e. they are the winding numbers of the two reduced 1D Hamiltonians
\begin{equation}
h_1(k_1)=h_0+h_2+(h_1+h_3) \exp(-ik_1)
\end{equation}
for $\nu_1$, and
\begin{equation}
h_2(k_2)=h_0+h_1+h_2 \exp(ik_2)+h_3 \exp(-ik_2)
\end{equation}
for $\nu_2$.
For example, assuming $h_0=h_1=h_3$ and $h_2/h_0=2$, the system is in the gapped (insulating) phase, i.e. $\nu_0=0$, and the winding numbers $\nu_{1,2}$ are given by 
$\nu_1=0$ and $\nu_2=1$. Figures 3(b) and (c) shows the numerically-computed evolution of the indices $\nu_{1,2}(\xi)$ versus propagation distance $\xi$, as obtained using Eq.(8) (i.e. neglecting GVM), in a 6-cm-long PPLN crystal with intensities $I_0=I_1=I_3=200$ MW/cm$^2$  and $I_2=4I_0=800$ MW/cm$^2$ of the four pump harmonics ($I_l \propto h_l^2 \; , l=0,1,2,3$).
Note the asymptotic convergence of $\nu_1(\xi)$ and $\nu_2(\xi) $ to the topological indices 0 and 1, respectively.\\
The above results suggest that SFG under a multi-frequency strong pump wave with incommensurate frequency comp'onents could provide a fascinating setup to emulate in photonics weak topological insulators in high dimensions.


\section {Conclusion} In conclusion, we unveiled that frequency conversion processes in nonlinear optical media, besides of obeying well-known conservation laws, are restricted by topological constraints and, alike topological insulators, can display topological phase transitions. We illustrated such a paradigm by considering sum frequency generation in second-order nonlinear media under a multi-frequency pump wave, showing that topological phase transitions can arise both at classical and quantum level. Our results shed new light on the foundations of nonlinear optics and the consequences of topological behaviors in nonlinear optics could be far-reaching for future applications of modern nonlinear and quantum optics.


\appendix
\begin{widetext}

\section{Topology with a time-periodic pump: classical analysis}
In this appendix we provide some technical details on the topological features of sum-frequency generation (SFG) when the strong pump wave is a periodic function of time. We use here a classical description of the frequency conversion  process using standard coupled-mode equations. In particular we discuss the effects of group velocity mismatch, which is not considered in the main text.

\subsection{\textbf{Coupled-mode equations}}
We assume that the nonlinear $\chi^{(2)}$ crystal is excited by a strong pump field, at carrier frequency $\omega_2$, and by a weak signal at carrier frequency $\omega_1$. In the undepleted pump approximation and after letting $\xi=z$ and $\eta=t-z/v_{g3}$ ($\xi$ describes the interaction distance in the crystal while $\eta$ is a retarded time in the reference frame of the pump wave), the evolution equations for the signal and SFG envelopes $\psi_{1,3}(\xi, \eta)$ are given by [Eqs.(3) and (4) in the main text]
\begin{eqnarray}
i \frac{\partial \psi_1}{\partial \xi}=i\left(  \frac{1}{v_{g2}}-\frac{1}{v_{g1}} \right) \frac{\partial \psi_1}{\partial \eta} +h(\eta) \psi_3 \\
i \frac{\partial \psi_3}{\partial \xi}=i\left(  \frac{1}{v_{g2}}-\frac{1}{v_{g3}} \right) \frac{\partial \psi_3}{\partial \eta} +h^*(\eta) \psi_1 
\end{eqnarray}
 where $h(\eta) \equiv - \sigma \psi_2^*(\eta)$ and $\psi_2(\eta)$ is the temporal profile of the undepleted pump envelope. The solution to Eqs.(A1) and (A2) cannot be given in an exact closed form rather generally (see \cite{r34b} and Sec.3 of Appendix A). However, when the group velocity mismatch between the waves is negligible, i.e. $v_{g1}=v_{g2}=v_{g3}$, the retarded time $\eta$ enters in the equations as a parameter, and the most general solution displays Rabi-like oscillations along the $\xi$ coordinate, i.e. oscillation cycles alternating SFG ($ \omega_1+ \omega_2 \rightarrow \omega_3$) and difference frequency generation ($ \omega_3- \omega_2 \rightarrow \omega_1$), namely one has
 \begin{equation}
 \left(
 \begin{array}{c}
 \psi_1( \xi, \eta) \\
 \psi_3(\xi, \eta)
 \end{array}
 \right)=
 \left(
 \begin{array}{cc}
\cos [\Delta (\eta) \xi ] &- i \sin[ \Delta (\eta) \xi] \exp[i \varphi(\eta)] \\
-i \sin[ \Delta (\eta) \xi] \exp[-i \varphi(\eta)] & \cos [\Delta (\eta) \xi ]
 \psi_3(\xi, \eta)
 \end{array}
 \right)=
 \left(
 \begin{array}{c}
 \psi_1( 0, \eta) \\
 \psi_3(0, \eta)
 \end{array}
 \right) 
 \end{equation}
 where we have set $h(\eta) \equiv \Delta (\eta) \exp[i \varphi(\eta)]$. Let us assume that the crystal is excited at the entrance plane by a monochromatic signal wave,  $\psi_1(\xi=0,\eta)$ independent of $\eta$, and $\psi_3(\xi=0, \eta)=0$. Assuming, without loss of generality, $\psi_1(\xi=0,\eta)=1$, one obtains
 \begin{equation}
 \psi_1(\xi,\eta)= \cos [\Delta(\eta) \xi] \; , \; \psi_3(\xi, \eta)=-i \sin [\Delta (\eta) \xi] \exp [-i \varphi(\eta)].
 \end{equation}
 \subsection{Calculation of the mean frequencies of signal and SFG waves}
 Let us assume that the pump wave is periodic in time with period $T= 2 \pi / \Omega$. After introduction of the scaled time $k=\eta \Omega$, the solutions $\psi_1(\xi, \eta)$ and $\psi_3(\xi,\eta)$, given by Eq.(A4), are periodic in $k$ with $2 \pi$ period, and can be therefore expanded in Fourier series with $\xi$-dependent coefficients, i.e.
 \begin{equation}
 \psi_1(\xi,k)=\sum_l a_l ( \xi) \exp(-ikl) \; , \; \;  \psi_3(\xi,k)=\sum_l b_l ( \xi) \exp(-ikl).
 \end{equation}
 Clearly, the spectral amplitude $|a_l(\xi)|^2$ is the (non-normalized) probability that, after an interaction distance $\xi$ in the nonlinear crystal, the signal photon has a frequency $\omega_1+l\Omega$. Likewise, $|b_l(\xi)|^2$ is the (non-normalized) probability that, after an interaction distance $\xi$, the SFG photon has a frequency $\omega_3+l\Omega$. The mean frequencies of signal and SFG waves are thus given by
 \begin{equation}
 \langle \omega_1 \rangle=\omega_1+\Omega \frac{\sum_l l |a_l(\xi)|^2}{\sum_l |a_l(\xi)|^2} \; ,\;\;\;   \langle \omega_3 \rangle=\omega_3+\Omega \frac{\sum_l l |b_l(\xi)|^2}{\sum_l |b_l(\xi)|^2}.
\end{equation}
 To calculate the series on the right hand sides of Eq.(A6), let us use the following property of Fourier series, that can be readily proven: for any given function $f(k)=R(k) \exp[-i \theta(k)]$, periodic in $k$ with $ 2 \pi$ period, after letting $f(k)=\sum_l f_l \exp(-ikl)$, one has
  \begin{eqnarray}
  \sum_l |f_l|^2 & = &  \frac{1}{2 \pi} \int_{-\pi}^{\pi} dk |f(k)|^2=\frac{1}{2 \pi} \int_{-\pi}^{\pi} dk R^2(k) \\
  \sum_l l |f_l|^2 & = &  \frac{i}{2 \pi} \int_{-\pi}^{\pi} dk f^*(k) \frac{df}{dk} =\frac{1}{2 \pi} \int_{-\pi}^{\pi} dk R^2(k) \frac{d \theta}{dk}.
  \end{eqnarray}
 From Eqs.(A4), (A6), (A7) and (A8) one then obtains
 \begin{equation}
 \langle \omega_1 \rangle=\omega_1 \; ,\;\;\;   \langle \omega_3 \rangle=\omega_3+\Omega \frac
{\int_{-\pi}^{\pi} dk \sin^2 [ \Delta(k) \xi] \frac{\partial \varphi}{\partial k}}{\int_{-\pi}^{\pi} dk \sin^2[ \Delta (k) \xi ] },
\end{equation} 
 i.e. $\langle \omega_1 \rangle=\omega_1 $ and $\langle \omega_3 \rangle=\omega_3+ \nu \Omega$, where $\nu$ is given by Eq.(5) in the main text.

\subsection{Effects of group velocity mismatch (GVM)}
Let us assume that the group velocities of signal and SFG waves are not exactly matched with the one of the pump wave. In this case the initial-value problem of Eqs.(A1) and (A2) can be solved rather generally using inverse scattering methods \cite{r34b}. When the pump wave is periodic in time with period $T= 2 \pi / \Omega$, we can however look for a solution to Eqs.(A1) and (A2) as a Fourier series in $\eta$, with $\xi$-dependent coefficients. After letting
\begin{equation}
\psi_1(\xi, k)=\sum_l a_l( \xi) \exp(-i kl) \; , \; \; \psi_3(\xi, k)=\sum_l b_l( \xi) \exp(-i kl) 
\end{equation}
with $k=\eta \Omega$, from Eqs.(A1), (A2) and (A10) one readily obtains
\begin{eqnarray}
i \frac{d a_l}{d \xi} & = &l \delta_1 a_l+\sum_{\rho} h_{\rho} b_{l-\rho} \\
i \frac{d b_l}{d \xi} & = &l \delta_3 b_l+\sum_{\rho} h^*_{\rho} a_{l+\rho} 
\end{eqnarray}
where we have set $\delta_1 \equiv \Omega(1/v_{g2}-1/v_{g1})$, $\delta_3 \equiv \Omega(1/v_{g2}-1/v_{g3})$ and $h(k) \equiv \sum_l h_l \exp(-ikl)$. Equations  (A11) and (A12) basically describe at the classical level the coupled signal and SFG spectral component dynamics on a synthetic binary lattice, discussed in the main text [Fig.1(b) and 2(c)], with initial excitation of the site $l=0$ of sublattice A (the injected monochromatic signal wave).  As it can be seen, the GVM (i.e. $ \delta_{1,3} \neq 0$) introduces uniform gradients in the two sublattices, which spoil out the discrete translation invariance of the lattice and are responsible for Bloch-Zener-type dynamics. However, when the spectral extent of the strong pump wave is sufficiently narrow, i.e. in the limit $\Omega \rightarrow 0$, the GVM terms can be neglected for not too long interaction lengths $\xi$ in the crystal. The strength of the pump wave is measured, for example, by its Fourier terms $h_0 \sim h_1$, and thus in the absence of the GVM the spreading in the lattice occurs at a speed of the order $\sim h_0$. After an interaction length $\xi$, the excitation has diffused to about $h_0 \xi$ sites in the lattice, so that GVM effects are negligible provided that $| \delta_{1,3} \xi h_0 | \ll |h_0|$, i.e. provided that the propagation length $\xi$ satisfies the condition
\begin{equation}
\xi \ll  \frac{1}{\Omega} {\rm{ min}}_{k=1,3} \left|  \frac{1}{v_{g2}}-\frac{1}{v_{gk}}  \right|^{-1}.
\end{equation}
To illustrate the effects of GVM, let us consider SFG in a periodically-poled lithium niobate (PPLN) crystal with a strong pump at the wavelength $\lambda_2 = 810$ nm and  a weak signal at $\lambda_1=1.55 \; \mu$m, as in the example discussed in the main text (Fig.2). The SFG wave corresponds to
$\lambda_3 =532$ nm. We assume extraordinary wave propagation, with a nonlinear coefficient $d_{33} \simeq 27$ pm/V. Phase matching is realized by a
first-order QPM grating ($7.38 \; \mu$m period), so that the effective nonlinear coefficient of the interaction is $d_{e}=(2/ \pi) d_{33}$.
The group velocities of signal, pump and SFG waves, as calculated using Sellmeier equations \cite{r39}, are $v_{g1}=0.4581 c_0$,
$v_{g2}=0.4422c_0$ and  $v_{g3}=0.4069c_0$, where $c_0$ is the speed of light in vacuum. For a frequency $\Omega= 2 \pi \times 1$ GHz, from Eq.(A13) it follows that GVM is negligible for propagation lengths satisfying the condition $\xi \ll 24$ cm. Since the typical lengths of as nonlinear crystal are smaller than 5-10 cm, neglecting GVM is a justified assumption. Clearly, GVM effects can become important as the strong pump wave is spectrally broadened. As an example, in Fig.4 we depict the numerically-computed evolution of the index $\nu$ versus interaction length $\xi$ for a bichromatic pump, carrying the intensities $I_0=400$ MW/cm$^2$, $I_1=800$ MW/cm$^2$ at the frequencies $\omega_2$ and $\omega_2+\Omega$, for a few increasing values of $\Omega$. Note that the quantization of $\nu$ is spoiled out at high frequencies $\Omega$ as a consequence of GVM.

\begin{figure*}[h]
	\includegraphics[width=0.9\linewidth]{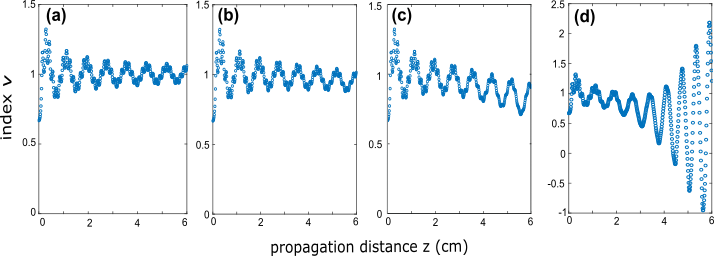}
\caption{\textbf{Effects of GVM.} Numerically-computed of the index $\nu$ versus propagation distance in the process of SFG in a PPLN crystal (crystal length 6 cm). The pump wave is bichromatic with pump intensities $I_0=400$ MW/cm$^2$, $I_1=800$ MW/cm$^2$ at frequencies $\omega_p$ and $\omega_p+\Omega$.  (a) $\Omega= 2 \pi \times 100$ MHz, (b) $\Omega= 2 \pi \times 1$ GHz, (c) $\Omega= 2 \pi \times 2$ GHz, and (d) $\Omega= 2 \pi \times 5$ GHz.}
\end{figure*}

\section{Topology with a time-periodic pump: quantum analysis}
\subsection{Heisenberg equations of motion and quantum walk on a synthetic frequency lattice}
The second-quantization Hamiltonian of the photon field in the nonlinear $\chi^{(2)}$ crystal under perfect phase matching and neglecting GVM is given by $\hat{H}=\hat{H}_0+\hat{H}_I$, where 
\begin{equation}
\hat{H}_0=\sum_n \hbar (\omega_1+n \Omega) \hat{a}^{\dag}_n \hat{a}_n+ \sum_n \hbar (\omega_3+n \Omega) \hat{b}^{\dag}_n \hat{b}_n+ \sum_n \hbar (\omega_2+n \Omega) \hat{c}^{\dag}_n \hat{c}_n
\end{equation}
is the Hamiltonian of the free photon field, and the trilinear Hamiltonian
\begin{equation}
\hat{H}_I=- \hbar \sigma v_{g} \sum_{n,l}(\hat{b}_{n+l} \hat{a}^{\dag}_{l} \hat{c}^{\dag}_{n}+H.c.)
\end{equation}
 is the interaction Hamiltonian. In the above equations $\hat{a}_n$, $\hat{a}^{\dag}_n$,
$\hat{b}_n$, $\hat{b}_n^{\dag}$ and $\hat{c}_n$, $\hat{c}^{\dag}_n$ are the annihilation and creation operators of the photon modes at frequencies $\omega_{1}+n \Omega$, $\omega_{3}+n \Omega$ and $\omega_{2}+n \Omega$, respectively, which satisfy the usual bosonic commutation relations, and $v_g=v_{g1}=v_{g2}=v_{g3}$ is the common group velocity of the three fields. The Heisenberg equations of motion of the destruction operators read
\begin{eqnarray}
i \frac{d \hat{a}_n}{dt}= \frac{1}{\hbar} [ \hat{a}_n,\hat{H} ]=(\omega_1+n\Omega) \hat{a}_n-\sigma v_{g} \sum_l\hat{c}_l^{\dag} \hat{b}_{n+l}\\
i \frac{d \hat{b}_n}{dt}= \frac{1}{\hbar} [ \hat{b}_n,\hat{H} ]=(\omega_3+n\Omega) \hat{a}_n-\sigma v_{g} \sum_l\hat{a}_l  \hat{c}_{n-l}\\
i \frac{d \hat{c}_n}{dt}= \frac{1}{\hbar} [ \hat{c}_n,\hat{H} ]=(\omega_2+n\Omega) \hat{a}_n-\sigma v_{g} \sum_l\hat{a}_l^{\dag} \hat{b}_{n+l} .
\end{eqnarray}
After the gauge transformation $\hat{a}_n \rightarrow \hat{a}_n \exp[-i(\omega_1+n \Omega) t]$, $\hat{b}_n \rightarrow \hat{b}_n \exp[-i(\omega_3+n \Omega) t]$ , 
and $\hat{c}_n \rightarrow \hat{c}_n \exp[-i(\omega_2+n \Omega) t]$, the above equations take the form
\begin{eqnarray}
i \frac{d \hat{a}_n}{dt}= -\sigma v_{g} \sum_l\hat{c}_l^{\dag} \hat{b}_{n+l}\\
i \frac{d \hat{b}_n}{dt}= -\sigma v_{g} \sum_l\hat{a}_l  \hat{c}_{n-l}\\
i \frac{d \hat{c}_n}{dt}= -\sigma v_{g} \sum_l\hat{a}_l^{\dag} \hat{b}_{n+l} 
\end{eqnarray}
where the interaction time $t$ is related to the propagation distance $z=\xi$ in the crystal by the relation
\begin{equation}
t=\xi / v_g.
\end{equation}
Assuming a strong and classical (coherent) pump wave, the operators $\hat{c}_l$ can be regarded as $c$-numbers, i.e. we can assume $\hat{c}_l \simeq \langle \hat{c}_l \rangle \equiv C_l$. In the undepleted pump approximation, such terms are constant and related to the incident pump wave profile $\psi_2(\eta)$ by the Fourier expansion
\begin{equation}
\psi_2(\eta) =\sum_l C_l \exp(-i l \Omega \eta).
\end{equation}
Therefore, for a strong classical pump and in the undepleted pump approximation, the Heisenberg equations for the destruction operators of signal and SFG photon fields read
\begin{eqnarray}
i \frac{d \hat{a}_n}{dt}=- \sigma v_{g} \sum_l C_l^* \hat{b}_{n+l}  \;, \;\; i \frac{d \hat{b}_n}{dt}=- \sigma v_{g} \sum_l C_l \hat{a}_{n-l} \;\;\;\;
\end{eqnarray}
which are Eqs.(6) given in the main text. Equations (B11) indicate that the signal and SFG photons undergo in tandem a continuous-time quantum walk on the sublattices A and B of a one-dimensional synthetic lattice in frequency space. In Bloch space, the Hamiltonian of the binary lattice is given by
\begin{equation}
H(k)=
\left( 
\begin{array}{cc}
0 & h(k) \\
h^*(k) & 0
\end{array}
\right)=\Delta(k) \cos [ \varphi(k)] \sigma_x-\Delta (k) \sin [ \varphi(k)] \sigma_y \;
\end{equation}
where we have set $h(k)=\Delta(k) \exp[i \varphi(k)]=-\sigma \psi_2^*(k)$  and where $\sigma_{x,y}$ are the Pauli matrices. The formal solution to Eq.(B11) can be written as
\begin{eqnarray}
\hat{a}_n(t) & = & \sum_l \left( \mathcal{A}_{n,l } \hat{a}_l(0)+ \mathcal{B }_{n,l} \hat{b}_l (0) \right) \\
\hat{b}_n(t) & = & \sum_l \left( \mathcal{C}_{n,l } \hat{a}_l(0)+ \mathcal{D }_{n,l} \hat{b}_l (0) \right).
\end{eqnarray}
where the $t$-dependent matrices $\mathcal{A}$, $\mathcal{B}$, $\mathcal{C}$ and $\mathcal{D}$ are determined by the propagator of the linear system and describe how an initial single-site excitation of the system, in either sublattice A or B, spreads in the lattice.

\subsection{Calculation of the mean frequency of signal and SFG photons} 
Let us assume that the crystal is excited at the entrance plane $\xi=0$ by a monochromatic signal field at frequency $\omega_1$ in an arbitrary quantum state, given by a superposition of Fock states, namely  let us assume
\begin{equation}
| \psi(0) \rangle= \sum_{l=1}^{\infty}  \frac{\alpha_l }{\sqrt{l !}} \hat{a}^{\dag l}_0 |0 \rangle
\end{equation}
 with arbitrary amplitudes $\alpha_l$ and $\sum_l | \alpha_l|^2=1$. Note that excitation with a single-photon Fock state corresponds to $\alpha_l=\delta_{l,1}$, whereas excitation with a classical field (a coherent state) corresponds to a Poisson distribution $\alpha_l=\alpha^l \exp(- |\alpha|^2/2) / \sqrt{l!}$, with $\alpha= \psi_1(0)$. The mean number of photons carried by the input signal wave is $\langle \psi(0)|  \hat{a}_0^{\dag}  \hat{a}_0  | \psi(0) \rangle=\sum_l  l | \alpha_l |^2$.
 
 After a propagation distance $\xi=v_{g} t$, the mean value of the frequency of the signal and SFG photon fields can be calculated as
\begin{eqnarray}
\langle \omega_1 \rangle & = & \omega_1+ \Omega \frac{\sum_n n \langle \psi(0) | \hat{a}_n^{\dag}(t)  \hat{a}_n(t)  | \psi(0) \rangle}{ \sum_n \langle \psi(0) | \hat{a}_n^{\dag}(t)  \hat{a}_n(t)  | \psi(0) \rangle} \\
\langle \omega_3 \rangle & = & \omega_3+ \Omega \frac{\sum_n n \langle \psi(0) |  \hat{b}_n^{\dag}(t) \hat{b}_n(t) | \psi(0) \rangle}{\sum_n \langle \psi(0) |\hat{b}_n^{\dag} (t) \hat{b}_n(t) | \psi(0) \rangle} 
\end{eqnarray}
The mean values entering in Eqs.(B16) and (B17) can be readily computed using Eqs.(B13), (B14) and (B15).  For example, one has
\begin{eqnarray}
\langle \psi(0) | \hat{a}_n^{\dag}(t) \hat{a}_n(t) | \psi(0) \rangle & = &  \sum_{\rho,\sigma} \langle \psi(0) |   \left( \mathcal{A}^*_{n, \sigma} \hat{a}^{\dag}_{\sigma}(0) + \mathcal{B}^{*}_{n, \sigma} \hat{b}^{\dag}_{\sigma}(0) \right) \left( \mathcal{A}_{n, \rho} \hat{a}_{\rho}(0) + \mathcal{B}_{n, \rho} \hat{b}_{\rho}(0) \right)  | \psi(0) \rangle \nonumber \\
& = &  \sum_{\rho} | \mathcal{A}_{n, \rho}|^2 \langle \psi(0) |  \hat{a}^{\dag}_{\rho} (0)  \hat{a}_{\rho} (0) | \psi(0) \rangle=|\mathcal{A}_{n, 0}|^2 \langle \psi(0) |  \hat{a}^{\dag}_{0} (0) \hat{a}_{0} (0)   | \psi(0) \rangle.
\end{eqnarray}
Similarly, one has
\begin{equation}
\langle \psi(0) | \hat{b}_n^{\dag}(t) \hat{b}_n(t) | \psi(0) \rangle = |\mathcal{C}_{n, 0}|^2 \langle \psi(0) | \hat{a}^{\dag}_{0} (0)  \hat{a}_{0}(0) | \psi(0) \rangle
\end{equation}
and thus
\begin{equation}
\langle \omega_1 \rangle = \omega_1+\Omega \frac{ \sum_n n |\mathcal{A}_{n,0}|^2}{\sum_n  |\mathcal{A}_{n,0}|^2} \;, \; \;  \langle \omega_3 \rangle = \omega_1+\Omega 
\frac{\sum_n n |\mathcal{C}_{n,0}|^2}{\sum_n  |\mathcal{C}_{n,0}|^2}.
\end{equation}
Equation (B20) clearly shows that the mean frequencies of signal and SFG photons do not depend on the initial quantum state $|\psi(0) \rangle$, and thus they should reproduce the result obtained by the classical analysis. In fact, the sums $\sum_n  |\mathcal{A}_{n,0}|^2$, $\sum_n n |\mathcal{A}_{n,0}|^2$, $\sum_n  |\mathcal{C}_{n,0}|^2$ and $\sum_n n |\mathcal{C}_{n,0}|^2$ entering in Eq.(B20) and associated to the quantum walk on the binary lattice with chiral symmetry can be calculated in terms of $h(k)=\Delta(k) \exp[i \varphi(k)]$ using the method described in Refs. \cite{r47,r49}, and read
\begin{equation}
\sum_n  |\mathcal{A}_{n,0}|^2= \frac{1}{2 \pi}   \int_{-\pi}^{\pi} dk \cos^2 [ \Delta(k) \xi] \; , \;\;\; \sum_n n |\mathcal{A}_{n,0}|^2=0 
\end{equation}
\begin{equation}
 \sum_n  |\mathcal{C}_{n,0}|^2= \frac{1}{2 \pi}   \int_{-\pi}^{\pi} dk \sin^2 [ \Delta(k) \xi] 
 \; ,\;\;\, \sum_n n |\mathcal{C}_{n,0}|^2= \frac{1}{2 \pi}  \int_{-\pi}^{\pi} dk \sin^2 [ \Delta(k) \xi] \left( \frac{d \varphi}{dk} \right).
\end{equation}
Therefore, one obtains $\langle \omega_1 \rangle= \omega_1$ and $\langle \omega_3 \rangle=\omega_3 + \nu \Omega$, where
\begin{equation}
\nu= \frac{\int_{-\pi}^{\pi} dk \sin^2 [ \Delta(k) \xi] \left( \frac{d \varphi}{dk} \right)}{\int_{-\pi}^{\pi} dk \sin^2 [ \Delta(k) \xi]}.
\end{equation}
Equation (B23) exactly reproduces the result obtained by the classical analysis (Sec.2 of Appendix A).

\section{Topology with a multifrequency aperiodic pump}
Let us assume that the strong pump wave envelope $\psi_2(\eta)$ is aperiodic in time and given by the superposition of $N$ mutually-incommensurate frequencies $\Omega_1$, $\Omega_2$,..., $\Omega_N$. In this case, the signal and SFG photons undergo a quantum walk on a high-dimensional synthetic lattice in frequency space \cite{r54}, which can display nontrivial topological features. For the sake of simplicity, we will consider the case of $N=2$ incommensurate frequencies $\Omega_1$ and $\Omega_2$, however the analysis can be readily extended to an arbitrary number of mutually incommensurate frequencies.
\subsection{Classical analysis}
In the classical analysis of SFG with two incommensurate frequencies $\Omega_1$ and $\Omega_2$ of the pump wave, it is worth introducing the two dimensionless variables $k_1=\Omega_1 \eta$ and $k_2=\Omega_2 \eta$, and considering the pump waveform $\psi_2(\eta)$ as a periodic function of the two independent variables $k_1$, $k_2$, i.e. $\psi_2=\psi_2(k_1,k_2)$. We can thus  expand $\psi_2(\eta)$ in double Fourier series as 
\begin{equation}
 \psi_2(k_1,k_2)=\sum_{n,m} C_{n,m} \exp(-ik_1n-ik_2m).
\end{equation} 
Neglecting group velocity mismatch effects and assuming a monochromatic injected signal field at the entrance of the crystal, the envelopes of signal and SFG waves at the propagation distance $\xi$ are given by
\begin{equation}
\psi_1(\xi,k_1,k_2)= \cos[\Delta(k_1,k_2) \xi] \; , \;\;\; \psi_3(\xi,k_1,k_2)=-i \sin [ \Delta (k_1,k_2) \xi ] \exp[-i \varphi(k_1,k_2)]
\end{equation}
where $h(k_1,k_2) \equiv - \sigma \psi_2^*(k_1,k_2)$ is written in terms of amplitude and phase as  $h(k_1,k_2) \equiv \Delta (k_1,k_2) \exp[i \varphi (k_1,k_2)]$.
Let us introduce the Fourier expansions for the two fields, with $\xi$-dependent coefficients, by letting
 \begin{equation}
 \psi_1(\xi,k_1,k_2)=\sum_{l,n} a_{l,n} ( \xi) \exp(-ik_1l-ik_2n) \; , \; \;  \psi_3(\xi,k_1,k_2)=\sum_{l,n} b_{l,n} ( \xi) \exp(-ik_1l-ik_2 n)
 \end{equation}
 Clearly, the spectral amplitude $|a_{l,n}(\xi)|^2$ is the (non-normalized) probability that, after an interaction distance $\xi$ in the nonlinear crystal, the signal photon has a frequency $\omega_1+l\Omega_1+n\Omega_2$. Likewise, $|b_{l,n}(\xi)|^2$ is the (non-normalized) probability that, after an interaction distance $\xi$, the SFG photon has a frequency $\omega_3+l\Omega_1+n\Omega_2$. The mean frequencies of signal and SFG waves are thus given by
 \begin{equation}
 \langle \omega_1 \rangle=\omega_1+ \frac{\sum_{l,n} (l \Omega_1+n \Omega_2) |a_{l,n}(\xi)|^2}{\sum_{l,n} |a_{l,n}(\xi)|^2} \; ,\;\;\;   \langle \omega_3 \rangle=\omega_3+\frac{\sum_{l,n} (l \Omega_1+n \Omega_2) |b_{l,n}(\xi)|^2}{\sum_{l,n} |b_{l,n}(\xi)|^2}
 \end{equation}
 To calculate the series on the right hand sides of Eq.(C4), let us use the following property of double Fourier series: for any two-dimensional function of the form $f(k_1,k_2)=\sum_{l.n} f_{l,n} \exp(-ik_1l-ik_2n)$, i.e. periodic in $k_1$ and $k_2$ with period $2 \pi$, after letting
 $f(k_1,k_2)=R(k_1,k_2) \exp[-i \theta(k_1,k_2)]$, one has
  \begin{eqnarray}
  \sum_{l,n} |f_{l,n}|^2 & = &  \frac{1}{(2 \pi)^2} \iint_{-\pi}^{\pi} dk_1dk_2 |f(k_1,k_2)|^2=\frac{1}{(2 \pi)^2} \iint_{-\pi}^{\pi} dk_1dk_2 R^2(k_1,k_2) \\
  \sum_{l,n} l |f_{l,n}|^2 & = &  \frac{i}{(2 \pi)^2} \iint_{-\pi}^{\pi} dk_1dk_2 f^*(k_1,k_2) \frac{\partial f}{\partial k_1} =\frac{1}{(2 \pi)^2} \iint_{-\pi}^{\pi} dk_1 dk_2 R^2(k_1,k_2) \frac{\partial \theta}{\partial k_1}\\
  \sum_{l,n} n |f_{l,n}|^2 & = &  \frac{i}{(2 \pi)^2} \iint_{-\pi}^{\pi} dk_1dk_2 f^*(k_1,k_2) \frac{\partial f}{\partial k_2} =\frac{1}{(2 \pi)^2} \iint_{-\pi}^{\pi} dk_1 dk_2 R^2(k_1,k_2) \frac{\partial \theta}{\partial k_2}.
  \end{eqnarray}
Using such identities, from Es.(C2), (C3) and Eq.(C4) one finally obtains
\begin{equation}
\langle \omega_1 \rangle= \omega_1 \; \; , \;\;\; \langle \omega_3 \rangle=\omega_3+\nu_1 \Omega_1+\nu_2 \Omega_2
\end{equation}
where we have set
\begin{equation}
\nu_{1,2}=\frac{\iint_{-\pi}^{\pi} dk_1 dk_2 \sin^2[\Delta(k_1,k_2) \xi] \left( \frac{\partial \varphi}{\partial k_{1,2}} \right) }{\iint_{-\pi}^{\pi} dk_1 dk_2 \sin^2[\Delta(k_1,k_2) \xi ]}.
\end{equation}

\subsection{Quantum analysis}
The second-quantization Hamiltonian of the photon field in the nonlinear $\chi^{(2)}$ crystal under perfect phase matching and neglecting GVM is given by $\hat{H}=\hat{H}_0+\hat{H}_I$, where 
\begin{equation}
\hat{H}_0=\sum_{n.m} \hbar (\omega_1+n \Omega_1+m \Omega_2) \hat{a}^{\dag}_{n,m} \hat{a}_{n,m}+ \sum_{n,m} \hbar (\omega_3+n \Omega_1+m \Omega_2) \hat{b}^{\dag}_{n,m} \hat{b}_{n,m}+ \sum_{n,m} \hbar (\omega_2+n \Omega_1+m \Omega_2) \hat{c}^{\dag}_{n,m} \hat{c}_{n,m}
\end{equation}
is the Hamiltonian of the free photon field, and
\begin{equation}
\hat{H}_I=- \hbar \sigma v_{g} \sum_{n_1,n_2,l_1,l_2}(\hat{b}_{n_1+l_1,n_2+l_2} \hat{a}^{\dag}_{l_1,l_2} \hat{c}^{\dag}_{n_1,n_2}+H.c.)
\end{equation}
 is the interaction Hamiltonian. In the above equations $\hat{a}_{n,m}$, $\hat{a}^{\dag}_{n.m}$,
$\hat{b}_{n,m}$, $\hat{b}_{n,m}^{\dag}$ and $\hat{c}_{n,m}$, $\hat{c}^{\dag}_{n,m}$ are the annihilation and creation operators of the photon modes at frequencies $\omega_{1}+n \Omega_1+m\Omega_2$, $\omega_{3}+n \Omega_1+m \Omega_2$ and $\omega_{2}+n \Omega_1+m\Omega_2$, respectively, which satisfy the usual bosonic commutation relations. Proceeding as in Sec.1 of Appendix B, assuming a strong and classical pump field and in the rotating-wave frame, the Heisenberg equations of motion of the destruction operators $\hat{a}_{n,m}$ and $\hat{b}_{n,m}$ read
\begin{eqnarray}
i \frac{d \hat{a}_{n,m}}{dt}=- \sigma v_{g} \sum_{l_1,l_2} C_{l_1,l_2} ^* \hat{b}_{n+l_1,m+l_2} , \; i \frac{d \hat{b}_{n,m}}{dt}=- \sigma v_{g} \sum_{l_1,l_2} C_{l_1,l_2} \hat{a}_{n-l_1,m-l_2} \;\;\;\;
\end{eqnarray}
where $C_{l_1,l_2}$ are the Fourier coefficients of the classical pump envelope $\psi_2(\eta)$, namely
\begin{equation}
\psi_2(\eta) =\sum_{l_1,l_2} C_{l_1,l_2} \exp(-i l_1 \Omega_1 \eta-i l_2 \Omega_2 \eta)
\end{equation}
and where in Eq.(C12) the interaction time $t$ is related to the interaction length $\xi$ in the crystal by the relation $t=\xi / v_g$. Let us assume that the crystal is excited at the entrance plane $\xi=0$ by a monochromatic signal field at frequency $\omega_1$ in an arbitrary quantum state, given by a superposition of Fock states, namely  let us assume
\begin{equation}
| \psi(0) \rangle= \sum_{l=1}^{\infty}  \frac{\alpha_l }{\sqrt{l !}} \hat{a}^{\dag l}_{0,0} |0 \rangle
\end{equation}
 with arbitrary amplitudes $\alpha_l$ and $\sum_l | \alpha_l|^2=1$. Proceeding as in Sec.2 of Appendix B, it can be shown that, after an interaction length $\xi$, the mean frequencies of signal and SFG photons do not depend on the initial quantum state $| \psi(0) \rangle$ and reproduce the classical result, given by Eqs.(C8) and (C9). 

\subsection{Topological properties}
Equations (C12) indicate that the signal and SFG photons undergo in tandem a continuous-time quantum walk on the sublattices A and B of a two-dimensional synthetic lattice in frequency space. In Bloch space, the Hamiltonian of the 2D lattice reads
\begin{equation}
H(k_1,k_2)=
\left( 
\begin{array}{cc}
0 & h(k_1,k_2) \\
h^*(k_1,k_2) & 0
\end{array}
\right)=\Delta(k_1,k_2) \cos [ \varphi(k_1.k_2)] \sigma_x- \Delta(k_1,k_2)  \sin [ \varphi(k_1,k_2)] \sigma_y \;
\end{equation}
where we have set $h(k_1,k_2)=\Delta(k_1,k_2) \exp[i \varphi(k_1,k_2)]=-\sigma \psi_2^*(k_1,k_2)$  and where $\sigma_{x,y,z}$ are the Pauli matrices.
 Such a synthetic lattice in frequency space provides a 2D extension of the SSH model introduced in recent works \cite{r55,r56,r57,r58} and represents an important example of a 2D weak topological insulator \cite{r57,r58} sustaining flat-band edge states \cite{r55,r56}. Here we briefly illustrate the topological properties of this model (for details see \cite{r57,r58}). The Hamiltonian $H(k_1,k_2) \equiv H(k)$ displays the following symmetries:\\
 (i) {\it Chiral symmetry}, namely $H(k) \sigma_z=-\sigma_z H(k)$.\\
 (ii) {\it $\mathcal{PT}$ symmetry}, namely $\mathcal{PT} H(k)=H(k) \mathcal{PT}$ with parity operator $\mathcal{P}=\sigma_x$ and time reversal operator $\mathcal{T}=\mathcal{K}$ ($\mathcal{K}$ is the element-wise complex conjugation operation).\\
 (iii) {\it Inversion symmetry.} Provided that the Fourier amplitudes $C_{n,m}$ of the pump wave are real, $H(k)$ also shows inversion symmetry, namely $H(-k) \mathcal{P}=\mathcal{P} H(k)$.\\
 For such a 2D weak topological insulator, the Berry curvature identically vanishes in the entire Brillouin zone and non-trivial topological phases can be identified by the strong $\mathbb{Z}_2$ index $\nu_0$ \cite{r59} and by two weak $\mathbb{Z}_2$ indices $\bar{\nu}_{1,2}$  \cite{r55,r56} or equivalently 
 by the  vectorized Zak phase $\boldsymbol{\theta}=(\theta_1,\theta_2)$ in 2D \cite{r55,r60,r61,r62}.   
 
The strong index $\nu_0$ is given by the relation
\begin{equation}
(-1)^{\nu_0}=\prod_{i=1
}^{4} \delta_i
\end{equation}
where $\delta_i=\pm 1$ is the parity eigenvalue of the Bloch eigenstates at the four time-reversal invariant momenta $(k_1,k_2)= \pi (n_1,n_2)$, with $n_{1,2}=0,1$. One has $\nu_0=0$ if and only if the two lattice bands are gapped, i.e. $h(k) \neq 0$ over the entire Brillouin zone: $\nu_0=0$ thus corresponds to an insulating phase \cite{r57}.\\

The 2D vectorized Zak phase is defined by \cite{r62}
 \begin{equation}
 \boldsymbol{\theta}=-\frac{1}{2 \pi} \iint_{-\pi}^{\pi} dk_1 dk_2 {\rm Tr}[ \mathbf{A} (k_1,k_2) ]
 \end{equation}
where $\mathbf{A}_{n,m}= \langle u_{n} | i \nabla_{\mathbf{k}} |u_{m} \rangle$ ($n,m= \pm$) is the Berry connection, $u_n(k_1,k_2)$ is the periodic part of the Bloch wave function in the $n$-th band, and the trace is taken over the  occupied bands of the lattice. For the Hamiltonian (C15), the periodic part of the Bloch functions in the two bands is given by
\begin{equation}
u_{\pm}(k_1,k_2) = \frac{1}{\sqrt{2}}
\left(
\begin{array}{c}
\exp[i \varphi (k_1,k_2)] \\
\pm 1
\end{array}
\right)
\end{equation}
so that one readily obtains $\mathbf{A}_{+,+}=\mathbf{A}_{-,-}=-(1/2) \nabla_{\mathbf{k}} \varphi$ and thus
\begin{equation}
 \boldsymbol{\theta}=\frac{1}{4 \pi} \iint_{-\pi}^{\pi} dk_1 dk_2 \nabla_{\mathbf{k}} \varphi .
\end{equation}
Note that, in the gapped phase ($\nu_0=0$), for any line $k_2$ the integral (winding number)
\[
\frac{1}{2 \pi} \int_{-\pi}^{\pi} dk_1 \frac{\partial \varphi}{\partial k_1}
\]
does not depend on $k_2$: in fact, it is an integer and its value cannot change as we adiabatically vary $k_2$, unless the gap closes. Likewise, the integral
\[
\frac{1}{2 \pi} \int_{-\pi}^{\pi} dk_2 \frac{\partial \varphi}{\partial k_2}
\]
does not depend on the line $k_1$ in the insulating phase. Therefore, when $\nu_0=0$ one obtains $\boldsymbol{\theta}=(\theta_1,\theta_2)$ for the 2D quantized Zak phase, with
\begin{equation}
\theta_{1,2}=\frac{1}{2} \int_{-\pi}^{\pi} dk_{1,2} \frac{\partial \varphi}{\partial k_{1,2}}.
\end{equation}
We remark that, since the Berry connection is gauge dependent, the Zak phase components $\theta_1$ and $\theta_2$ are uniquely defined mod $ 2 \pi$, i.e. they can uniquely take the two possible values $0$ and $ \pi$.
The vectorized Zak phase can be readily associated to the indices $\nu_{1,2}$, given by Eq.(C9) and determining the mean frequency of the SHG photons. In fact, in the topological insulating phase $\Delta(k_1,k_2)$ does not vanish in the entire Brillouin zone, and thus for large enough interaction length $\xi$ we may set $\sin^2 [ \Delta(k_1,k_2) \xi] \simeq 1/2$ in Eq.(C9), yielding
\begin{equation}
\nu_{1,2} \simeq \frac{1}{4 \pi^2} \iint_{-\pi}^{\pi} dk_1 dk_2 \frac{\partial \varphi}{\partial k_{1,2}}= \frac{1}{2 \pi} \int_{-\pi}^{\pi} dk_{1,2} \frac{\partial \varphi}{\partial k_{1,2}}.
\end{equation}
A comparison of Eqs.(C20) and (C21) yields
\begin{equation}
\theta_{1,2}= \pi \nu_{1,2} \; \; \;  \; ({\rm mod \;} 2 \pi ).
\end{equation}
As an illustrative example, let us assume 
\begin{equation}
h(k_1,k_2)=h_0+h_1 \exp(-ik_1)+h_2 \exp(ik_2)+h_3 \exp(-ik_1-ik_2),
\end{equation} 
which corresponds to a pump envelope $\psi_2(\eta)= -(1/ \sigma) h^*(k_1,k_2)$ comprising the four frequencies $\omega_2$, $\omega_2 -\Omega_1$, $\omega_2+\Omega_2$, and $\omega_2-\Omega_1-\Omega_2$ with amplitudes $h_0$, $h_1$, $h_2$ and $h_3$, respectively. The value of the strong  topological  index $\nu_0$ can be computed from the parity eigenvalue of the Bloch eigenstates at the four time-reversal invariant momenta $(k_1,k_2)= \pi (n_1,n_2)$  ($n_{1,2}=0,1$) according to Eq.(C16), and reads
\begin{equation}
(-1)^{\nu_o}= {\rm{sign}} \left\{ (h_0+h_1+h_2+h_3)(h_0-h_1+h_2-h_3)(h_0+h_1-h_2-h_3)(h_0-h_1-h_2+h_3)  \right\}.
\end{equation}
In the insulating phase, i.e. for $\nu_0=0$, the winding numbers $\nu_{1,2}$ can be calculated from Eq.(C21) along the lines $k_{2,1}=0$, i.e. they are the winding numbers of the two reduced 1D Hamiltonians
\begin{equation}
h_1(k_1)=h_0+h_2+(h_1+h_3) \exp(-ik_1)
\end{equation}
for $\nu_1$, and
\begin{equation}
h_2(k_2)=h_0+h_1+h_2 \exp(ik_2)+h_3 \exp(-ik_2)
\end{equation}
for $\nu_2$. The 2D weak topological insulator associated to this model is illustrated in Fig.3(a) of the main text. Depending on the values of the pump amplitudes $h_0$, $h_1$, $h_2$ and $h_3$, different topological phases, corresponding to different values of the topological numbers, can be obtained.
\end{widetext}


\begin{thebibliography}{31}

\bibitem{r1}
P. A. Franken, A. E. Hill, C.W. Peters, and G. Weinreich, Generation of Optical Harmonics,
Phys. Rev. Lett. {\bf 7}, 118 (1961).

\bibitem{r2}
J. A. Armstrong, N. Bloembergen, J. Ducuing, and P. S. Pershan, Interactions between Light Waves in a Nonlinear Dielectric,
Phys. Rev. {\bf 127}, 1918 (1962).

\bibitem{r3}
R. W. Boyd, {\it Nonlinear Optics}, 3rd ed. (Elsevier, 2008).

\bibitem{r4}
M.M. Fejer, Nonlinear optical frequency conversion, Phys. Today {\bf 47},  25 (1994).

\bibitem{r4b}
S.M. Saltiel, A.A. Sukhorukov, and Y.S. Kivshar,
Multistep Parametric Processes in Nonlinear Optics, Progress in Optics {\bf 47}, 1 (2005).

\bibitem{r5}
M. H. Dunn and M. Ebrahimzadeh, Parametric Generation of Tunable Light from Continuous-Wave to Femtosecond Pulses, Science {\bf 286}, 1513
(1999).

\bibitem{r6}
S. Kim, J. Jin, Y. J. Kim, I. Y. Park, Y. Kim, and
S. W. Kim, High-harmonic generation by resonant plasmon
field enhancement, Nature {\bf 453}, 757 (2008).

\bibitem{r7}
R. W. Andrews, R. W. Peterson, T. P. Purdy, K. Cicak, R. W. Simmonds, C. A. Regal, and K. W. Lehnert, Bidirectional and efficient conversion between microwave and optical light, Nature Phys. {\bf 10}, 321 (2014).
 
\bibitem{r8}
G. Cerullo and S. De Silverstri,
Ultrafast optical parametric amplifiers,
Rev. Sci. Instr. {\bf 74}, 1 (2003).

\bibitem{r9}
R. Thomson, C. Leburn, and D. Reid, eds., {\it Ultrafast Nonlinear Optics} (Springer, Berlin, 2013).

\bibitem{r10}
M. Geissbuehler, L. Bonacina, V. Shcheslavskiy, N. L. Bocchio, S. Geissbuehler, M. Leutenegger, I. Marki, J. P. Wolf, and T. Lasser, Nonlinear correlation spectroscopy, Nano Lett. {\bf 12}, 1668 (2012).

\bibitem{r11}
R. Slusher, L. Hollberg, B. Yurke, J. Mertz, and J. Valley, Observation of Squeezed States Generated by Four-Wave Mixing in an Optical Cavity, Phys. Rev. Lett. {\bf 55}, 2409 (1985).

\bibitem{r12}
H.J. Kimble, Squeezed states of light: an (incomplete) survey of experimental progress and prospects, Phys. Rep.
{\bf 219}, 227 (1992).

\bibitem{r13}
J. Huang and P. Kumar, Observation of Quantum Frequency Conversion, Phys. Rev. Lett. {\bf 68}, 2153 (1992).

\bibitem{r14}
S. Barz, G. Cronenberg, A. Zeilinger, and P. Walther,
Heralded generation of entangled photon pairs, Nature
Photon. {\bf 4}, 553 (2010).

\bibitem{r15}
Y.-Z. Sun, Y.-P. Huang, and P. Kumar, Photonic Nonlinearities via Quantum Zeno Blockade,
Phys. Rev. Lett. {\bf 110}, 223901 (2013).

\bibitem{r15b}
T. Guerreiro, E. Pomarico, B. Sanguinetti, N. Sangouard, J. S. Pelc, C. Langrock, M. M. Fejer, H. Zbinden, R. T. Thew, and N. Gisin, 
Interaction of independent single photons based
on integrated nonlinear optics,
Nature Commun. {\bf 4}, 2324 (2013).

\bibitem{r16}
Helge R\"utz, K.-H. Luo, H. Suche, and C. Silberhorn, Quantum Frequency Conversion between Infrared and Ultraviolet,
Phys. Rev. Applied {\bf 7}, 024021 (2017).

\bibitem{r17}
Y. Guo, P. P. Ho, H. Savage, D. Harris, P. Sacks, S.
Schantz, F. Liu, N. Zhadin, and R. R. Alfano, Second-harmonic tomography of tissues, Opt. Lett.
{\bf 22}, 1323 (1997).

\bibitem{r18}
P. Pantazis, J. Maloney, D. Wu, and S. E. Fraser, Second harmonic generating (SHG) nanoprobes for in vivo imaging, Proc. Natl. Acad. Sci. {\bf 107}, 14535  (2010).




\bibitem{r19}
H. Suchowski,G. Porat, and A. Arie, 
Adiabatic processes in frequency conversion, Laser \& Photon. Rev. {\bf 8}, 333 (2013).

\bibitem{r20}
D. Smirnova, D. Leykam, Y. Chong, and Y. Kivshar, 
Nonlinear topological photonics,
Appl. Phys. Rev. {\bf 7}, 021306 (2021).

\bibitem{r21}
A. Karnieli, Y. Li, and A. Arie, The geometric phase in nonlinear frequency conversion, Front. Phys. {\bf 17}, 12301 (2022).

\bibitem{r21b}
S. Longhi, Zitterbewegung of optical pulses in nonlinear frequency conversion,
J. Phys. B: At. Mol. Opt. Phys. {\bf 43} 205402 (2010).

\bibitem{r22}
N. V. Bloch, K. Shemer, A. Shapira, R. Shiloh, I. Juwiler, and A. Arie, Twisting light by nonlinear photonic crystals, Phys. Rev. Lett. {\bf 108}, 233902 (2012).

\bibitem{r23}
V. Peano, M. Houde, F. Marquardt, and A.A. Clerk, Topological Quantum Fluctuations and Traveling Wave Amplifiers,
Phys. Rev. X {\bf 6}, 041026 (2016).

\bibitem{r24}
S. Longhi, Transparency in nonlinear frequency conversion, Phys. Rev. A {\bf 93}, 043822 (2016).

\bibitem{r24b}
S. Longhi, Third-harmonic generation in quasi-phase-matched $\chi^{(2)}$ media with missing second harmonic, 
Opt. Lett. {\bf 32}, 1791 (2007).


\bibitem{r25}
A. Karnieli and A. Arie,
All-Optical Stern-Gerlach Effect,
Phys. Rev. Lett. {\bf 120}, 053901 (2018).

\bibitem{r26}
W. Zhang, J. Tang, Y. Ming, C. Zhang, and Y. Lu,
Optical-field topological phase transition in nonlinear frequency conversion,
Opt. Express  {\bf 28}, 2818 (2020) 

\bibitem{r27}
O. Yesharim, A. Karnieli, S. Jackel, G. Di Domenico, S. Trajtenberg-Mills, and A. Arie,
 Observation of the all-optical Stern-Gerlach effect in nonlinear optics, Nat. Photon. {\bf 16}, 582 (2022).


\bibitem{r28}
L. Lu, J. D. Joannopoulos, and M. Soljacic, Topological photonics, Nat. Photonics {\bf 8}, 821 (2014).   

\bibitem{r29}
 L. Lu, J. D. Joannopoulos, and M. Soljacic, Topological states in photonic systems, Nat. Phys. {\bf 12}, 626 (2016).   
 
 \bibitem{r30}
L. Yuan, Q. Lin, M. Xiao, and S. Fan, Synthetic dimension in photonics, Optica {\bf 5}, 1396 (2018).

\bibitem{r31}
T. Ozawa, H. M. Price, A. Amo, N. Goldman, M. Hafezi, L. Lu, M. C. Rechtsman, D. Schuster, J. Simon, O. Zilberberg, and I. Carusotto, Topological photonics, Rev. Mod. Phys. 
{\bf 91}, 015006 (2019).

\bibitem{r32}
E. Lustig and M. Segev,
Topological photonics in synthetic dimensions, Adv. Opt. and Photon. {\bf 13}, 426 (2021) 

\bibitem{r33}
T. Morimoto and N. Nagaosa,
Topological nature of nonlinear optical
effects in solids, Sci. Adv. {\bf 2}, e1501524 (2016).

\bibitem{r34}
N. Bloemberger, Conservation laws in nonlinear optics, J. Opt. Soc. Am. {\bf 70}, 1429 (1980).

\bibitem{r34b}
D. J. Kaup, A. Reiman, and A. Bers, Space-time evolution of nonlinear three-wave interactions. I. Interaction in a homogeneous medium,
Rev. Mod. Phys. {\bf 51}, 275 (1979).

\bibitem{r35}
K. Dholakia, N. B. Simpson, M. J. Padgett, and L. Allen, Second-harmonic generation and the orbital angular momentum of light,
Phys. Rev. A {\bf 54}, R3742 (1996).


\bibitem{r45}
J. K. Asboth, L. Oroszlany, and A. Palyi, A Short Course on Topological Insulators: Band-structure, topology and edge states in one and two dimensions
(Springer, Lecture Notes in Physics vol. {\bf 919}, 2016).

\bibitem{r46b}
X.-L. Qi and S.C. Zhang, Topological insulators and superconductors. Rev. Mod.
Phys. {\bf 83}, 1057 (2011).

\bibitem{r46c}
C.-K. Chiu, J.C.Y. Teo, A.P. Schnyder, and S. Ryu, Classification of topological quantum matter with symmetries,
Rev. Mod. Phys. {\bf 88}, 035005 (2016).

\bibitem{r36}
P. Kumar,  Quantum frequency conversion, Opt. Lett. {\bf 15}, 1476 (1990).

\bibitem{r38}
D.S. Hum and M.M. Fejer, Quasi-phasematching, C. R. Physique {\bf 8}, 180 (2007).


\bibitem{r39}
G. J. Edwards and M. Lawrence, A temperature-dependent dispersion equation for congruently grown lithium niobate, Opt. Quantum Electron. {\bf 16},
373 (1984).

\bibitem{r40}
D.F. Walls and R. Barakat, Quantum-Mechanical Amplification and Frequency Conversion with a Trilinear Hamiltonian,
Phys. Rev. A {\bf 1}, 446 (1970).

\bibitem{r41}
B. Dayan, Theory of two-photon interactions with broadband down-converted light and entangled photons,
Phys. Rev. A {\bf 76}, 043813 (2007).

\bibitem{r42}
S. Blum, G.A. Olivares-Renteria, C. Ottaviani, C. Becher, and G. Morigi,
Single-photon frequency conversion in nonlinear crystals, Phys. Rev. A {\bf 88}, 053807 (2013).

\bibitem{r43}
A. Christ, B. Brecht, W. Mauerer, and C. Silberhorn,
Theory of quantum frequency conversion and type-II parametric down-conversion in the high-gain regime,
New J. Phys. {\bf 15}, 053038 (2013).

\bibitem{r44}
J.M. Donohue, M.D. Mazurek, and K.J. Resch, Theory of high-efficiency sum-frequency generation for single-photon waveform conversion,
Phys. Rev. A {\bf 91}, 033809  (2015).


\bibitem{r46}
B. Perez-Gonzalez, M. Bello, A. Gomez-Leon, and G. Platero, 
Interplay between long-range hopping and disorder in topological systems,
Phys. Rev. B {\bf 99}, 035146 (2019).

\bibitem{r47}
F. Cardano, A. D$^{\prime}$Errico, A. Dauphin, M. Maffei, B. Piccirillo, C. de
Lisio, G. De Filippis, V. Cataudella, E. Santamato, L. Marrucci, M.
Lewenstein, and P. Massignan, Detection of Zak phases and topological invariants in a
chiral quantum walk of twisted photons, Nat. Commun. {\bf 8}, 15516 (2017).

\bibitem{r48}
M. Maffei, A. Dauphin, F. Cardano, M. Lewenstein, and P. Massignan, Topological
characterization of chiral models through their long time dynamics, New J. Phys.
{\bf 20}, 013023 (2018).

\bibitem{r49}
S. Longhi, Probing one-dimensional topological phases in
waveguide lattices with broken chiral symmetry, Opt. Lett. {\bf 43}, 4639 (2018).

\bibitem{r50}
Y. Wang, Y.-H. Lu, F. Mei, J. Gao, Z.-M. Li, H. Tang, S.-L. Zhu, S. Jia, and X.-M. Jin,
Direct Observation of Topology from Single-Photon Dynamics,
Phys. Rev. Lett. {\bf 122}, 193903 (2019).

\bibitem{r51}
D. Xie, W. Gou, T. Xiao, B. Gadway, and B. Yan,
Topological characterizations of an extended
Su-Schrieffer-Heeger model, npj Quantum Inf. {\bf 55}, (2019).


\bibitem{r52}
Z.-Q. Jiao, S. Longhi, X.-W. Wang, J. Gao, W.-H. Zhou, Y. Wang, Y.-X. Fu,
L. Wang, R.-J. Ren, L.-F. Qiao, and X.-M. Jin,
Experimentally Detecting Quantized Zak Phases without
Chiral Symmetry in Photonic Lattices, Phys. Rev. Lett. {\bf 127}, 147401 (2021).

\bibitem{r53}
B.-H. Chen and D.-W. Chiou, An elementary rigorous proof of bulk-boundary correspondence in the generalized Su-Schrieffer-Heeger model, Phys. Lett. A {\bf 384}, 126168 (2020).




\bibitem{r54}
I. Martin, G. Refael, and B. Halperin, Topological Frequency Conversion in Strongly Driven Quantum Systems
Phys. Rev. X {\bf 7}, 041008 (2017).



\bibitem{r55}
L. Zhu, E. Prodan, and K. H. Ahn, Flat energy bands within antiphase and twin boundaries and at open edges in topological materials, Phys. Rev. B{\bf  99}, 041117(R) (2019).

\bibitem{r56}
K. Qian, L. Zhu, K.H. Ahn, and C. Prodan, Observation of Flat Frequency Bands at Open Edges and Antiphase Boundary Seams in Topological Mechanical Metamaterials,
Phys. Rev. Lett. {\bf 125}, 225501 (2020).

\bibitem{r57}
S. Jeon and Y. Kim, Two-dimensional weak-type insulators in inversion-symmetric crystals,
Phys. Rev. B {\bf 105}, L121101 (2022).

\bibitem{r58}
H. Yang, L. Song, Y. Cao, and P. Yan, Experimental Realization of Two-Dimensional Weak Topological
Insulators, Nano Lett. {\bf 22}, 3125 (2022).



\bibitem{r59}
L. Fu and C.L. Kane, Topological insulators with inversion symmetry, Phys. Rev. B {\bf 76}, 045302 (2007).


\bibitem{r60}
G. van Miert, C. Ortix, and C. Morais Smith,
Topological origin of edge states in two-dimensional inversionsymmetric
insulators and semimetals, 2D Mater. {\bf 4}, 015023 (2017).

\bibitem{r61}
F. Liu and K. Wakabayashi,
Novel Topological Phase with a Zero Berry Curvature, Phys. Rev. Lett. {\bf 118}, 076803 (2017).


\bibitem{r62}
M. Kim and J. Rho,
Topological edge and corner states in a two dimensional
photonic Su-Schrieffer-Heeger lattice, Nanophoton. {\bf 9},  3227 (2020).


\end{thebibliography}
\end{document}